\documentstyle[14lomcon,cite,graphicx,bm]{article}

\begin{document}

\title{FLAVOR OSCILLATIONS OF LOW ENERGY NEUTRINOS \\ IN THE ROTATING NEUTRON STAR}

\author{Maxim Dvornikov \footnote{e-mail: maxim.dvornikov@usm.cl}}

\address{Departamento de F\'{i}sica y Centro de Estudios
Subat\'{o}micos, \\
Universidad T\'{e}cnica Federico Santa Mar\'{i}a, Casilla 110-V,
Valpara\'{i}so, Chile and \\
IZMIRAN, 142190, Troitsk, Moscow region, Russia}


\maketitle\abstracts{We study flavor oscillations of low energy
neutrinos propagating in dense matter of a rotating neutron star.
On the basis of the exact solutions of the wave equations for
neutrinos mass eigenstates we derive the transition probability
for neutrinos having big initial angular momentum. It is found
that flavor oscillations of neutrinos with energies of several
electron-Volts can be resonancely enhanced.}

It is known that neutrinos play a significant role at the last
stages of the evolution of massive stars. For example, almost 99\%
of the gravitational energy of a protoneutron star is carried away
during the supernova explosion. The remaining dense, compact
object, a neutron star, can have extreme properties: central
density $\sim 10^{14}\thinspace{\rm g/cc}$, magnetic field $\sim
10^{15}\thinspace{\rm G}$ and angular velocity $\sim
10^3\thinspace{\rm s}^{-1}$. In this short note we examine the
influence of the neutron star rotation on flavor oscillations of
neutrinos. Note that the propagation of neutrinos in rotating
matter was also studied in Refs.~\cite{StuROT,DvoROT}.

Let us study the evolution of the two flavor neutrinos system
$(\nu_\alpha,\nu_\beta)$ interacting with the background matter by
means of the electroweak forces. The Lagrangian for this system
has the form,
\begin{equation}\label{Lagrnu}
  {\cal L}=
  \sum_{\lambda=\alpha,\beta}
  \bar{\nu}_\lambda
  ({\rm i}\gamma^\mu\partial_\mu - f^\mu_\lambda\gamma_\mu^{\rm L})
  \nu_\lambda-
  \sum_{\lambda\lambda'=\alpha,\beta}
  m_{\lambda\lambda'} \bar{\nu}_\lambda \nu_{\lambda'},
\end{equation}
where $\gamma_\mu^{\rm L}=\gamma_\mu(1-\gamma^5)/2$. Supposing
that matter is electroneutral and all the background fermions
rotate as a rigid body we can express the external fields
$f^\mu_\lambda$ for $\alpha = \mu$ or $\tau$ and $\beta = e$
oscillations channel as,
\begin{equation}\label{falphabeta}
  f_\alpha^\mu = -\frac{G_{\rm F}}{\sqrt{2}}j_n^\mu,
  \quad
  f_\beta^\mu = \frac{G_{\rm F}}{\sqrt{2}}(2 j_e^\mu-j_n^\mu),
  \quad
  j_{e,n}^\mu = (n_{e,n}, n_{e,n}{\bf v}),
\end{equation}
where $G_{\rm F}$ is the Fermi constant, $n_{e,n}$ is the number
density of electrons and neutrons and ${\bf v}=(\bm{\Omega} \times
{\bf r})$ is the velocity of the background matter.

To study the evolution of the system~(\ref{Lagrnu}) we should
introduce the neutrino mass eigenstates $\psi_a$ to diagonalize
the mass matrix $(m_{\lambda\lambda'})$, ${\nu_{\lambda} =
(\exp[-{\rm i}\sigma_2 \theta])_{\lambda a}} \\ \times \psi_a$,
where $\theta$ is the vacuum mixing angle. We suggest that the
mass eigenstates are Dirac particles. In the basis of the mass
eigenstates $\psi_a$ neutrinos have definite masses $m_a$.

In the limit of small neutrino masses the wave equations for the
upper $\xi_a$ and lower $\eta_a$ chiral components of the spinor
$\psi_a^{\rm T}=(\xi_a,\eta_a)$ decouple. Therefore using
cylindrical coordinates $(r,\phi,z)$ with $\bm{\Omega} = \Omega
{\bf e}_z$ we can write the general expression for the two
component wave function $\eta_a$ in the form (see
Ref.~\cite{DvoROT}),
\begin{eqnarray}\label{gensol}
  \eta_a(r,\phi,t)=
  \sum_{\mathrm{n},s=0}^{\infty}
  \Big(
    a_{{\rm n}s}^{(a)}(t)\  u_{a,{\rm n}s}^{+{}}(r,\phi)
    \exp[-{\rm i}E_{\rm n}^{(a)+{}} t]
    \nonumber
    \\
    \qquad {}+
    b_{{\rm n}s}^{(a)}(t)\  u_{a,{\rm n}s}^{-{}}(r,\phi)
    \exp[-{\rm i}E_{\rm n}^{(a)-{}} t]
  \Big),
\end{eqnarray}
where the energy levels
\begin{equation}\label{enlevMP}
  E_{\rm n}^{(a)\pm{}} = - V_a \pm \sqrt{4V_a \Omega {\rm n} + m_a^2},
  \quad
  {\rm n}=0, 1, 2,\ldots
\end{equation}
have the discrete values and the basis spinors
\begin{equation}\label{basspinMP}
  u_{a,{\rm n}s}^{(\pm)}(r,\phi) =
  \sqrt{\frac{V_a\Omega}{2\pi}}
  \left(
  \begin{array}{c}
    I_{{\rm n}-1,s}(\rho_a) e^{{\rm i}(l-1)\phi} \\
    \mp {\rm i} I_{{\rm n},s}(\rho_a) e^{{\rm i}l\phi} \
  \end{array}
  \right),
  \quad l = {\rm n}-s,
\end{equation}
are expressed in terms of the Laguerre functions $I_{{\rm
n},s}(\rho_a)$ of the dimensionless argument $\rho_a = V_a \Omega
r^2$, $V_1=G_{\rm F}(n_n-2n_e\sin^2\theta)/\sqrt{2}$ and
$V_2=G_{\rm F}(n_n-2n_e\cos^2\theta)/\sqrt{2}$ are the potentials
of the interaction of mass eigenstates with background matter.
Note that in Eqs.~(\ref{gensol})-(\ref{basspinMP}) we study
neutrinos propagating in the equatorial plane with $z=0$.

In Ref.~\cite{DvoROT} we obtained the general differential
equations for the coefficients $a_{{\rm n}s}^{(a)}(t)$ and
discussed the situation of the small initial angular momentum: $l
\ll s$. Now we study neutrino flavor oscillations for $l \gg s$,
i.e. particles with big initial angular momentum. Using the
results of our work~\cite{DvoROT} we get that in this situation
the differential equations for different $l$ and $s$ decouple and
we can describe the evolution of the system with help of the
single Schr\"{o}dinger equation,
\begin{equation}\label{Schreq}
  {\rm i}\frac{{\rm d}}{{\rm d}t}
  \left(
  \begin{array}{c}
    {\tilde a}_{l}^{(1)}\\
    {\tilde a}_{l}^{(2)}
  \end{array}
  \right)
  =
  \left(
  \begin{array}{cc}
    \omega/2 & \Delta \\
    \Delta & -\omega/2 \
  \end{array}
  \right)
  \left(
  \begin{array}{c}
   {\tilde a}_{l}^{(1)}\\
   {\tilde a}_{l}^{(2)}
  \end{array}
  \right),
\end{equation}
where the components $\tilde a_{l}^{(a)}$ of the ``wave function",
which now can be enumerated with the single quantum number ``$l$"
are related to the coefficients in Eq.~(\ref{gensol}) by the
formula, ${\tilde a}_{l}^{(a)} = (\exp[-{\rm i} \sigma_3
\omega/2])^a_b a_{l}^{(b)}$. The parameters of the effective
Hamiltonian in Eq.~(\ref{Schreq}) have the form,
\begin{equation}\label{paramSchreq}
  \Delta =
  \left(
    \frac{G_{\rm F}}{\sqrt{2}} -
    \frac{k}{2n_n}
  \right) n_e \sin2\theta,
  \quad
  \omega = \frac{\delta m^2}{2k},
  \quad
  k = \sqrt{4 V \Omega l},
\end{equation}
where $k$ is the effective momentum of neutrinos and $V = G_{\rm
F}n_n/\sqrt{2}$.

\begin{figure}
  \centering
  \includegraphics[scale=.39]{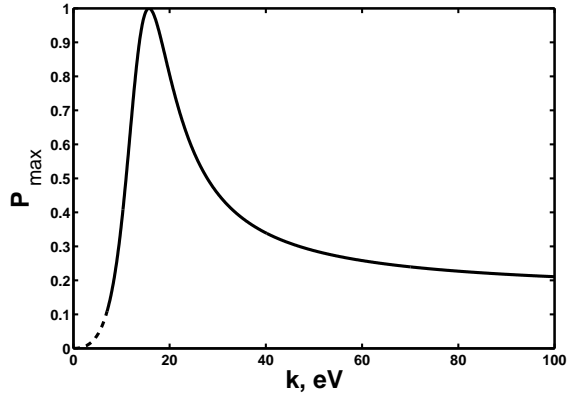}
  \caption{\label{fig1}
  The dependence of the maximal transition probability on the neutrino
  energy for the case $l \gg s$. This plot corresponds to $\nu_e \to
  \nu_\mu$ oscillations, with $\delta m^2 \approx 8.1\times
  10^{-5}\thinspace{\rm eV}^2$ and $\sin^2\theta \approx 0.3$ 
  and matter with $n_n = 10^{38}\thinspace{\rm cm}^{-3}$ and
  $Y_e = n_e/n_n = 3 \times 10^{-3}$. The very low energy part of the curve,
  which cannot be treated in frames of the quantum mechanical approach,
  is shown by the dashed line.}
\end{figure}

Using Eqs.~(\ref{Schreq}) and~(\ref{paramSchreq}) we can obtain
the transition probability in the form,
\begin{equation}\label{Ptr}
  P_{\beta \to \alpha}(x)=
  \frac{(\Delta \cos 2\theta + \omega \sin 2\theta/2)^2}
  {\Delta^2+(\omega/2)^2}
  \sin^2
  \left(
    \sqrt{\Delta^2+(\omega/2)^2}x
  \right).
\end{equation}
Let us discuss the oscillation scheme $\nu_e \to \nu_\mu$. In
Fig.~\ref{fig1} we present the maximal transition probability as a
function of the neutrino energy, built on the basis of
Eq.~(\ref{Ptr}).
Note that for the nuclear matter in $\beta$ equilibrium the number
density of electrons has the following value: $n_e \approx 3 \pi^2
n_n^2/(2m_n)^3$. As one can see from this picture, the transition
probability has a resonance behaviour. The maximal transition
probability is reached at $\sim 16\thinspace{\rm eV}$. At very
large energies the transition probability approaches to the limit
$P_{\rm max} \to \cos^2(2\theta) \approx 0.16$, a result that can
also be inferred from Eq.~(\ref{Ptr}). We should notice that the
solution presented in Eq.~(\ref{Ptr}) is not valid for very large
neutrino energies, because in that case the condition $l \gg s$ is
violated.

\section*{Acknowledgments}

The work has been supported by Conicyt (Chile), Programa
Bicentenario PSD-91-2006. The author is thankful to C.~O.~Dib and
for helpful discussions and the organizers of $14^{\rm
th}$~Lomonosov conference for the invitation.

\end{document}